\documentclass[aps,prl,twocolumn]{revtex4-1}
\usepackage{graphicx}
\usepackage{hyperref}
\usepackage{amsmath}
\usepackage{amssymb}
\usepackage{subfigure}
\usepackage{cancel}

\newcommand{\beq}{\begin{equation}}
\newcommand{\eeq}{\end{equation}}
\newcommand{\bea}{\begin{eqnarray}}
\newcommand{\eea}{\end{eqnarray}}

\begin{document}
\begin{flushright}
%\today \\
%%
UMD-PP-013-014\\
\end{flushright}
\title{Exploring a Dark Sector Through the Higgs Portal at a Lepton Collider}
%\today
 \author{Zackaria Chacko}
 \affiliation{Maryland Center for Fundamental Physics, Department of Physics, University of Maryland, College Park, MD 20742, USA}
\author{Yanou Cui}
\affiliation{Maryland Center for Fundamental Physics, Department of Physics, University of Maryland, College Park, MD 20742, USA}
\author{Sungwoo Hong}
\affiliation{Maryland Center for Fundamental Physics, Department of Physics, University of Maryland, College Park, MD 20742, USA}
%\preprint{UMD-PP-xxxxxx}

\begin{abstract}
    
We investigate the prospects for detecting a hidden sector at an $e^+ e^-$
collider. The hidden sector is assumed to be composed of invisible 
particles that carry no charges under the Standard Model gauge 
interactions, and whose primary interactions with ordinary matter are 
through the Higgs portal. We consider both the cases when the decays of an 
on-shell Higgs into a pair of hidden sector particles are kinematically 
allowed, and the case when such decays are kinematically forbidden. We 
find that at collider energies below a TeV, the most sensitive 
channel involves production of an on-shell or off-shell Higgs in 
association with a Z boson, and the subsequent decay of the Higgs into invisible
hidden sector states. Focusing on this channel, we find that with order 
a thousand inverse fb of data at 250 GeV, the decay branching fraction of an on-shell 
Higgs to invisible hidden sector states can be constrained to lie below half a percent. The 
corresponding limits on Higgs portal dark matter will be stronger than 
the bounds from current and upcoming direct detection experiments in much of 
parameter space. With the same amount of data at 500 GeV, assuming order one couplings, 
decays of an off-shell Higgs to hidden sector states with a total mass up to about 
200 GeV can also be probed. Both the on-shell and off-shell cases represent a 
significant improvement in sensitivity when compared to the Large Hadron Collider (LHC).

\end{abstract}


\maketitle

\section{Introduction} 

The existence of a hidden sector that does not transform under the 
Standard Model (SM) gauge interactions is one of the most intriguing 
possibilities for physics beyond the SM. Since the Higgs bilinear 
$H^{\dagger} H$ is the lowest dimension gauge invariant operator in the 
SM, the simplest possibility is that it is through this ``Higgs portal'' 
that the hidden sector primarily communicates with the 
SM~\cite{Patt:2006fw}. Such a hidden sector has been invoked to resolve 
many of the outstanding problems of the SM, including the observed 
abundance of dark matter (DM) 
\cite{Silveira:1985rk,McDonald:1993ex,Burgess:2000yq,Kim:2008pp,LopezHonorez:2012kv}. 
A hidden mirror sector that communicates with the SM through the Higgs 
portal~\cite{Foot1991bp,Foot:1991py} can also explain the stability of 
the electroweak 
scale~\cite{Chacko:2005pe,Barbieri:2005ri,Chacko:2005vw}. It is 
therefore important to understand the extent to which current and future 
collider programs will be able to probe this scenario.

The hidden sector states produced through the Higgs portal may be 
invisible at colliders, or may decay back into SM states. The latter 
case has been studied in, for 
instance,~\cite{Schabinger:2005ei,Strassler:2006im,Bowen:2007ia,Barger:2007im}. 
The focus of our work is on the invisible case. If the hidden sector 
states are sufficiently light, the Higgs can decay into them. Such 
invisible decays of the Higgs can be searched for at colliders. The 
current $2\sigma$ limit on the invisible branching fraction of the Higgs 
from the LHC is $\sim70\%$ \cite{ATLAS:2013pma,Chasco:2013pwa}, 
from the search of Z+Higgs with Higgs decaying to invisible states. It is 
expected that with 3000 fb$^{-1}$ of data at 14 TeV this limit can be 
improved to 
$\sim10\%$~\cite{Frederiksen:1994me,Eboli:2000ze,Ghosh:2012ep,Godbole:2003it,Bai:2011wz,Djouadi:2012zc,Peskin:2012we,ATLAS:2013hta,CMS:2013xfa}. 
A lepton collider would be able to improve this bound by more than an 
order of magnitude~\cite{snowmass}. Since some fraction of the produced 
Higgs particles now decay invisibly, there is also a uniform reduction 
in the rate to visible SM final states that can be used to set tighter 
limits on this scenario, both at the 
LHC~\cite{Bock:2010nz,Englert:2011yb,Englert:2011aa} and at lepton 
colliders, for example ~\cite{Walker:2013hka}. The current limit from global fitting with 8 TeV LHC data is 
$\sim20\%$ \cite{Belanger:2013kya, Djouadi:2013qya}.

If, however, the hidden sector states that the Higgs couples to are 
heavy, so that an on-shell Higgs cannot decay into them, the situation 
is much more challenging. Such states can now only be accessed through 
decays of an off-shell Higgs, and so the production cross section 
receives additional phase space suppression, without any corresponding 
reduction in the background. As a consequence, the LHC has only very 
limited sensitivity to such a scenario~\cite{Kanemura:2010sh}. Lepton 
colliders are expected to have somewhat better reach. It has been shown 
that a 5 TeV linear collider will be able to probe hidden sector states 
with a total mass up to about 200 GeV through the off-shell Higgs 
portal~\cite{Kanemura:2011nm}, assuming order one couplings. However, 
this analysis makes use of off-shell Higgs production through the Z 
boson fusion channel, which is highly suppressed at lower center of mass 
(CM) energies. The situation at lower CM energies is less well 
understood. A preliminary study~\cite{Matsumoto:2010bh} suggests that 
the reach of a 300 GeV linear collider is limited to hidden sector 
states with a total mass up to about 140 GeV, only slightly above the 
Higgs mass of 125 GeV. Given the importance of the question, a more 
thorough analysis is clearly warranted.

In this paper we investigate the prospects for detecting a hidden sector 
through the Higgs portal at a next generation $e^+ e^-$ collider, such 
as ILC or TLEP~\cite{Baer:2013cma,Gomez-Ceballos:2013zzn}. Accordingly, 
we focus on the proposed CM energies: 250 GeV, 350 GeV, 500 GeV and 1 
TeV at the ILC, and 240 GeV or 350 GeV at TLEP. For each case we conduct 
a parton-level study, with signals and backgrounds generated by Madgraph 
5~\cite{Alwall:2011uj}. In our analysis, we consider both the cases when 
decays of an on-shell Higgs into a pair of hidden sector particles are 
kinematically allowed, and the case when such decays are kinematically 
forbidden. We find that at these energies the most sensitive channel 
involves production of an on-shell or off-shell Higgs in association 
with a Z boson, and the subsequent invisible decay of the Higgs into 
hidden sector states. Focusing on this channel, we find that with 
$O(1000)\rm ~fb^{-1}$ of data at $250$ GeV, the branching fraction of an 
on-shell Higgs to invisible hidden sector states can be constrained to 
lie below half a percent. The corresponding limits on Higgs portal DM 
will be stronger than the bounds from current and upcoming direct 
detection experiments in a large part of parameter space. With the same amount 
of data at higher energies, assuming $O(1)$ couplings, decays of an 
off-shell Higgs to hidden sector states with a total mass up to about 
200 GeV can also be probed. Both the on-shell and off-shell cases 
represent a significant improvement in sensitivity when compared to 
the LHC. Earlier studies of the connection between dark matter direct detection
 and invisible Higgs decay at the LHC have been performed in \cite{Arbey:2013aba,Belanger:2013pna}, in the context of supersymmetric theories.

\section{Model}
We consider the simplest Higgs portal model where the hidden sector consists of a singlet scalar $\phi$ that couples to $H^{\dagger} H$ through renormalizable interactions~\cite{Burgess:2000yq}. We focus on the scenario where $\phi$ is stable, and may constitute the dark matter. Accordingly, we impose a $Z_2$ parity under which $\phi$ is odd while the SM fields are even. The Lagrangian involving $\phi$ is given by:
\beq
 \mathcal{L}=\mathcal{L}_{\rm SM}+\frac{1}{2}\partial^\mu\phi\partial_\mu\phi-\frac{1}{2}M_S^2\phi^2-\frac{c_S}{2}|H|^2\phi^2-\frac{\lambda}{4!}\phi^4,\label{lagrangian}
\eeq 
where $M_S$ is the bare mass of $\phi$. After electroweak symmetry breaking, $H$ gets vacuum expectation value $\langle H\rangle=(0,v)^T/\sqrt{2}$, where $v=246$ GeV. The physical mass of $\phi$ is then given by:
\beq
m_\phi^2=M_S^2+\frac{c_S v^2}{2}.
\eeq
We will parametrize the model by $(m_\phi, c_S)$ in our analysis.

\section{Signals and backgrounds}
  
\begin{figure}
      {\includegraphics[height=20mm,width=35mm]{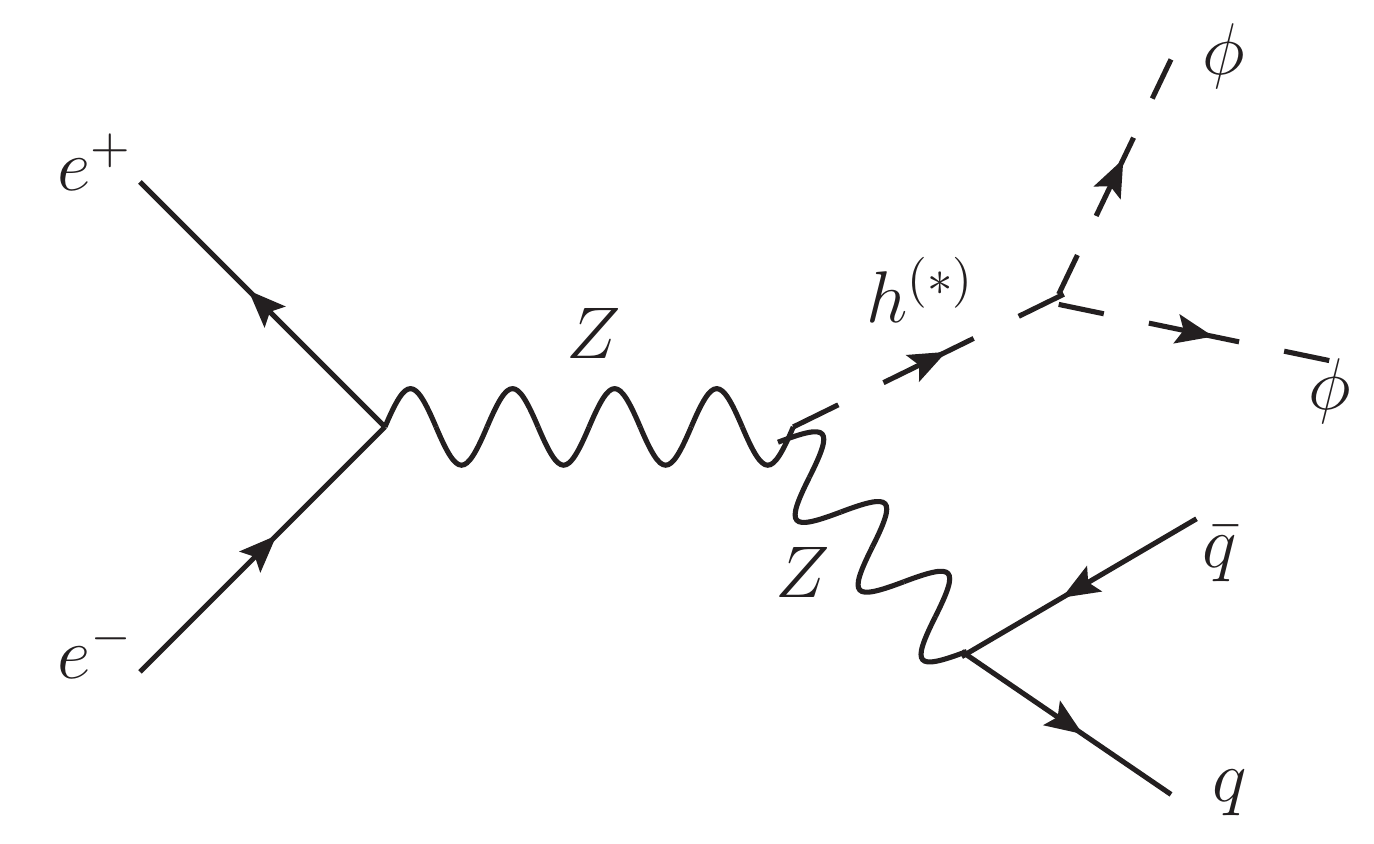}}
         \mbox{ 
   \subfigure {\includegraphics[height=20mm,width=35mm]{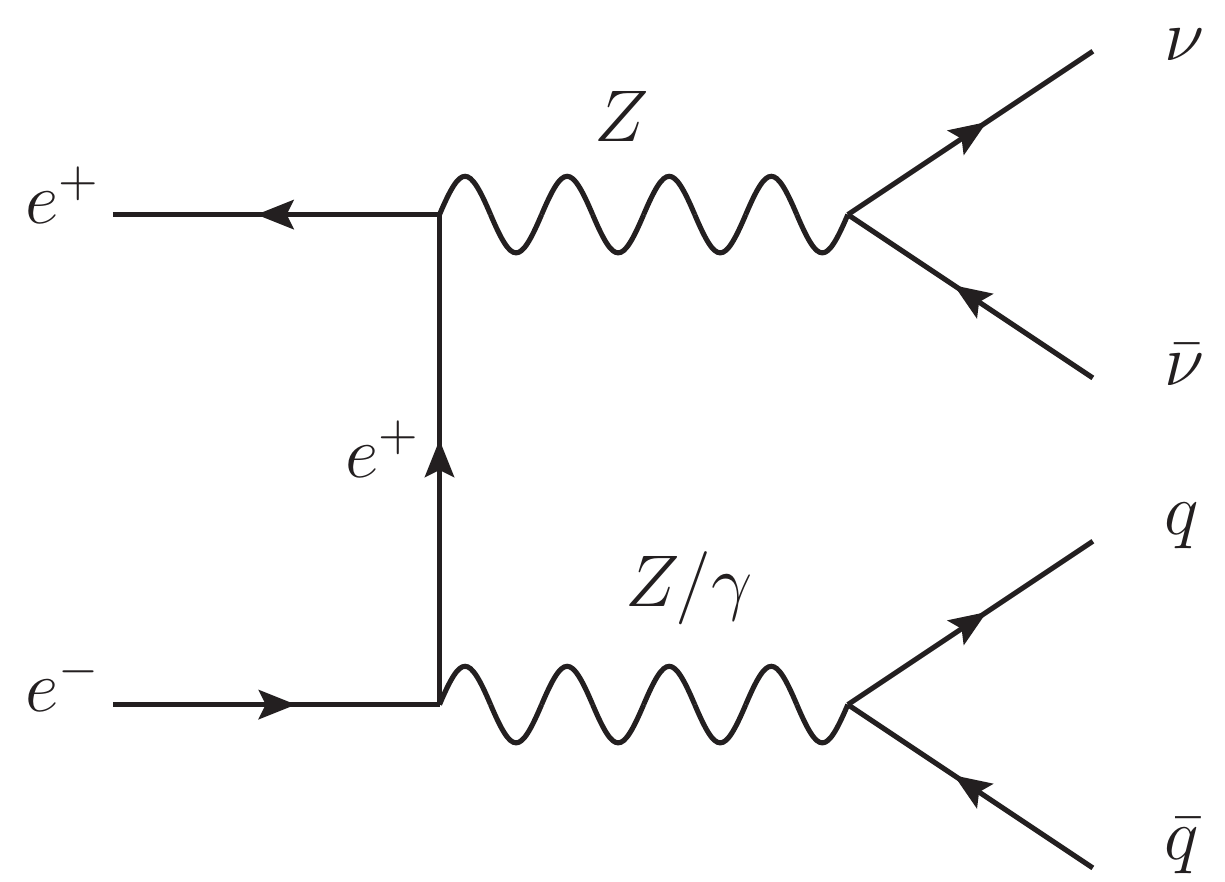}}
    \subfigure{\includegraphics[height=20mm,width=35mm]{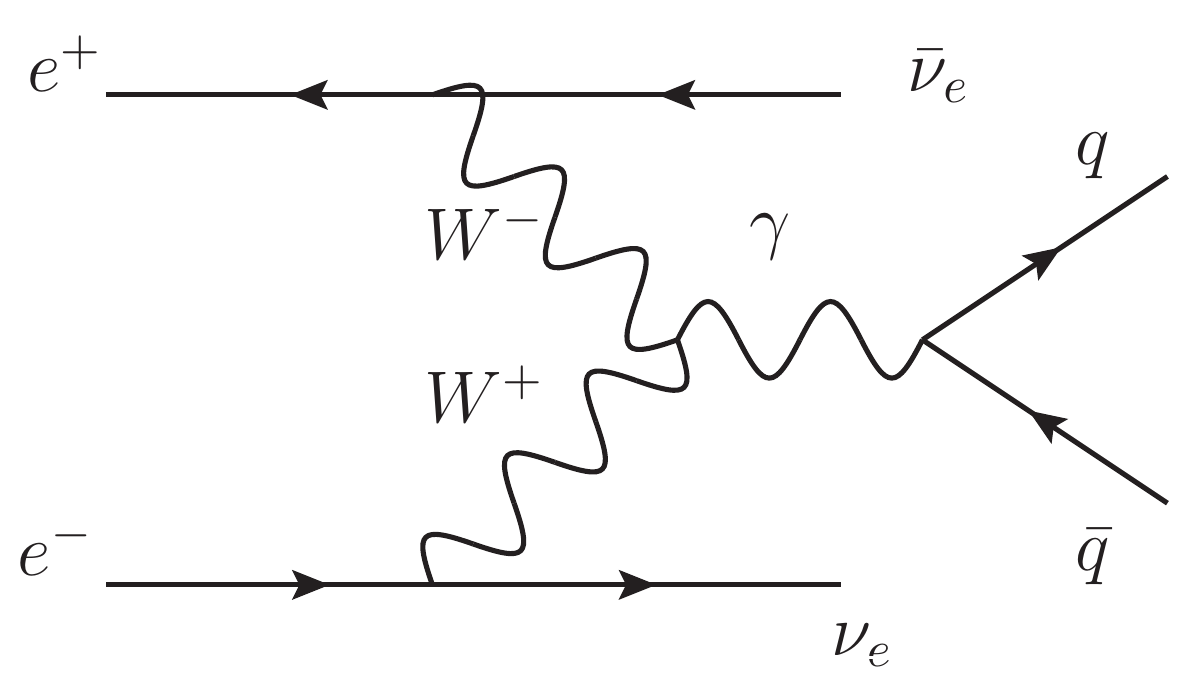}}  
      }
  \caption{Associate production (AP) channel: signal process (upper) and examples of leading background (lower).}\label{fig:AP_Fdiagrams}
  \end{figure}

\begin{figure}
      {\includegraphics[height=20mm,width=35mm]{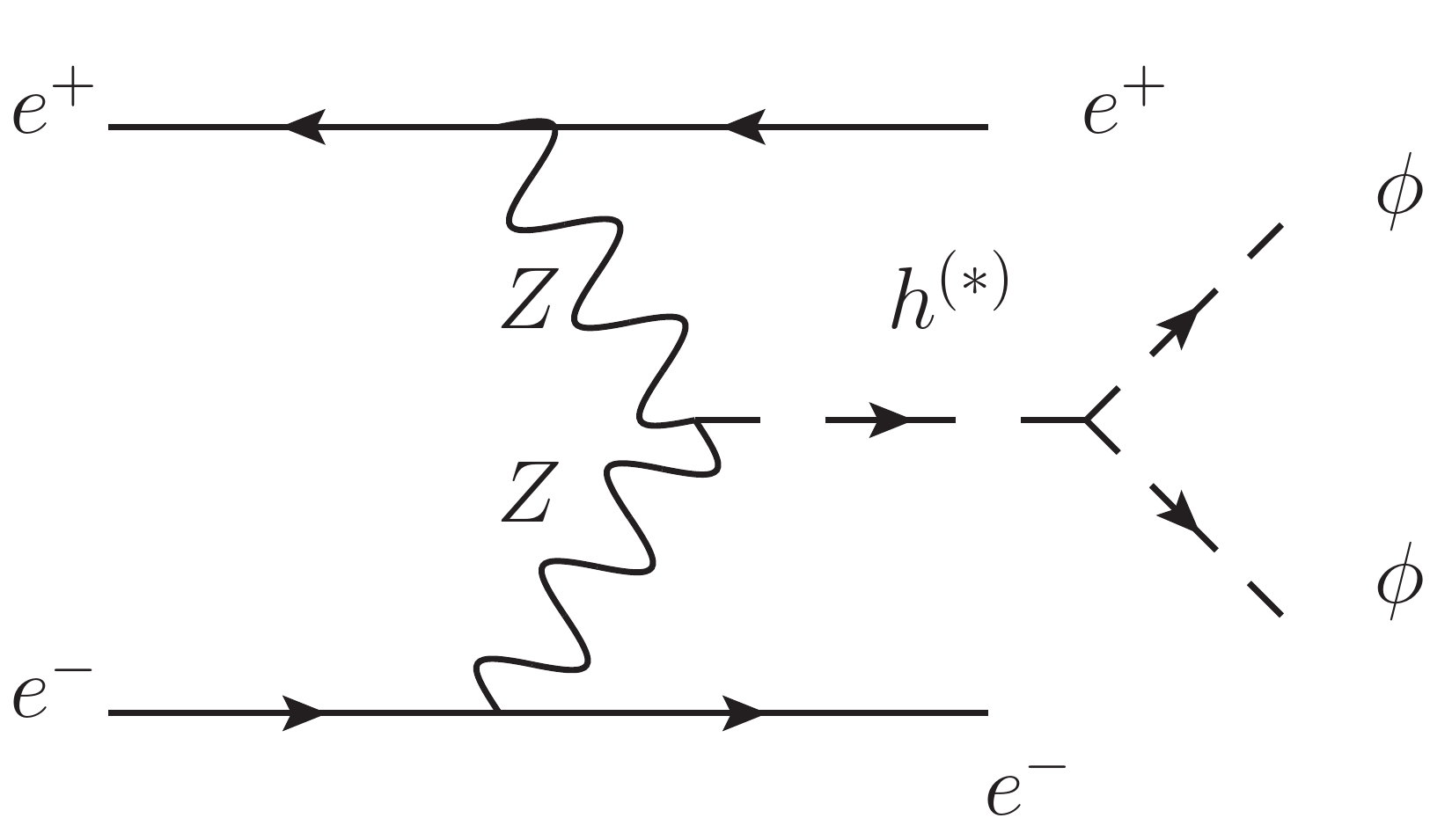}}
         \mbox{ 
   \subfigure {\includegraphics[height=20mm,width=35mm]{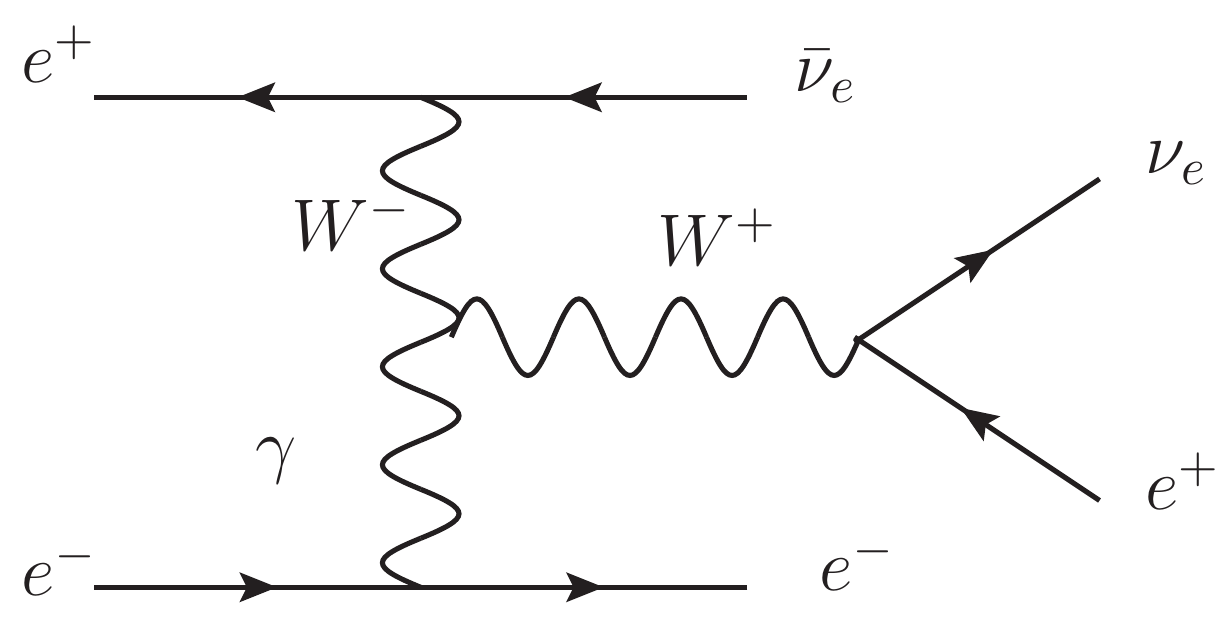}}
    \subfigure{\includegraphics[height=20mm,width=35mm]{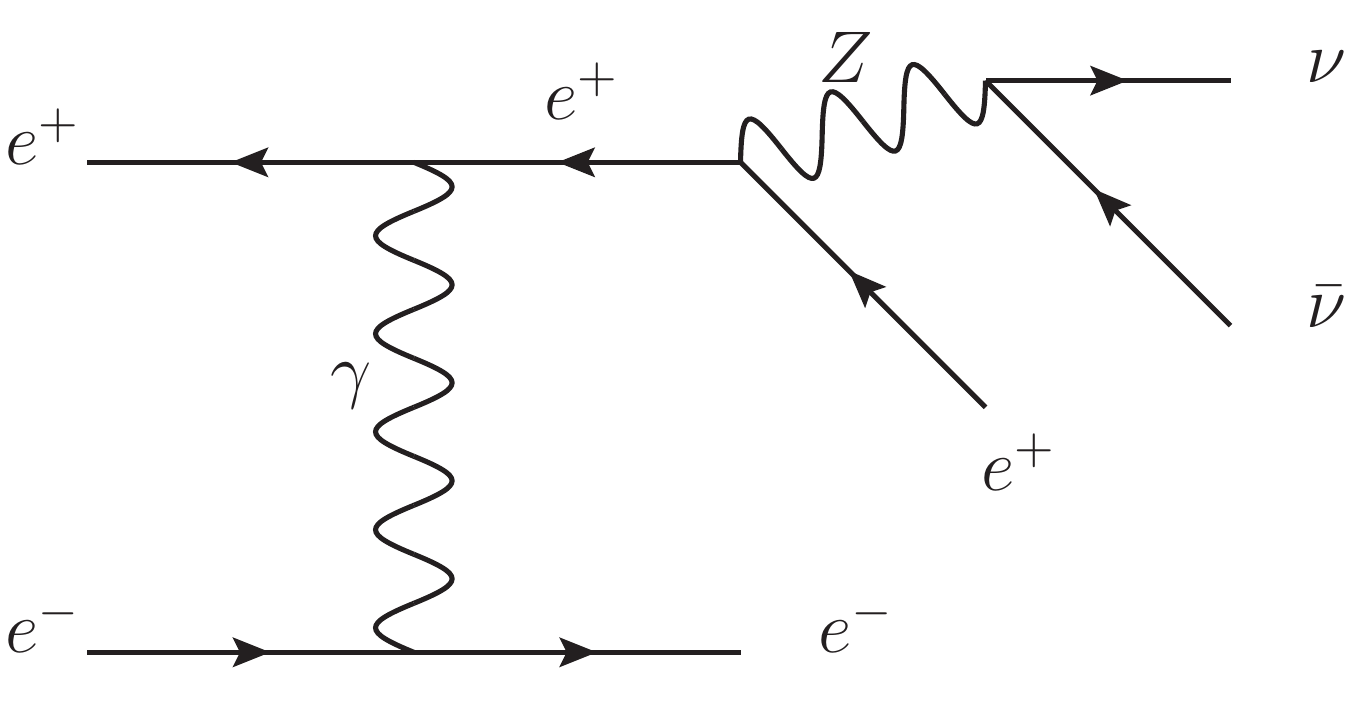}}  
      }
  \caption{Z-fusion (ZF) production channel: signal process (upper) and examples of leading background (lower).}\label{fig:ZF_Fdiagrams}
  \end{figure}

At an $e^+e^-$ collider, there are two different channels through which 
$\phi$ can be pair produced. One channel involves associated production 
(AP) with $Z$-bremsstrahlung $e^+e^- \rightarrow Zh^{(*)}\rightarrow 
Z\phi\phi$. The $Z$ subsequently decays, primarily hadronically, so that 
the corresponding signal involves jets and missing energy. The other 
production channel involves $Z$-fusion (ZF), $e^+e^-\rightarrow 
e^+e^-h^{(*)}\rightarrow e^+e^-\phi\phi$. The corresponding signal 
involves an electron-positron pair and missing energy. When 
$m_\phi<\frac{m_H}{2}$, the $\phi$'s in these events can be produced 
from on-shell Higgs decays, and contribute to the invisible width of the 
Higgs. However, when $m_\phi>\frac{m_H}{2}$ the production has to go 
through off-shell Higgs, and the search is more challenging. The AP and 
ZF signal processes and examples of major backgrounds are shown in 
Fig.(\ref{fig:AP_Fdiagrams}) and Fig.(\ref{fig:ZF_Fdiagrams}).

\begin{figure}
   %\mbox{   
   \subfigure   {\includegraphics[height=40mm,width=60mm]{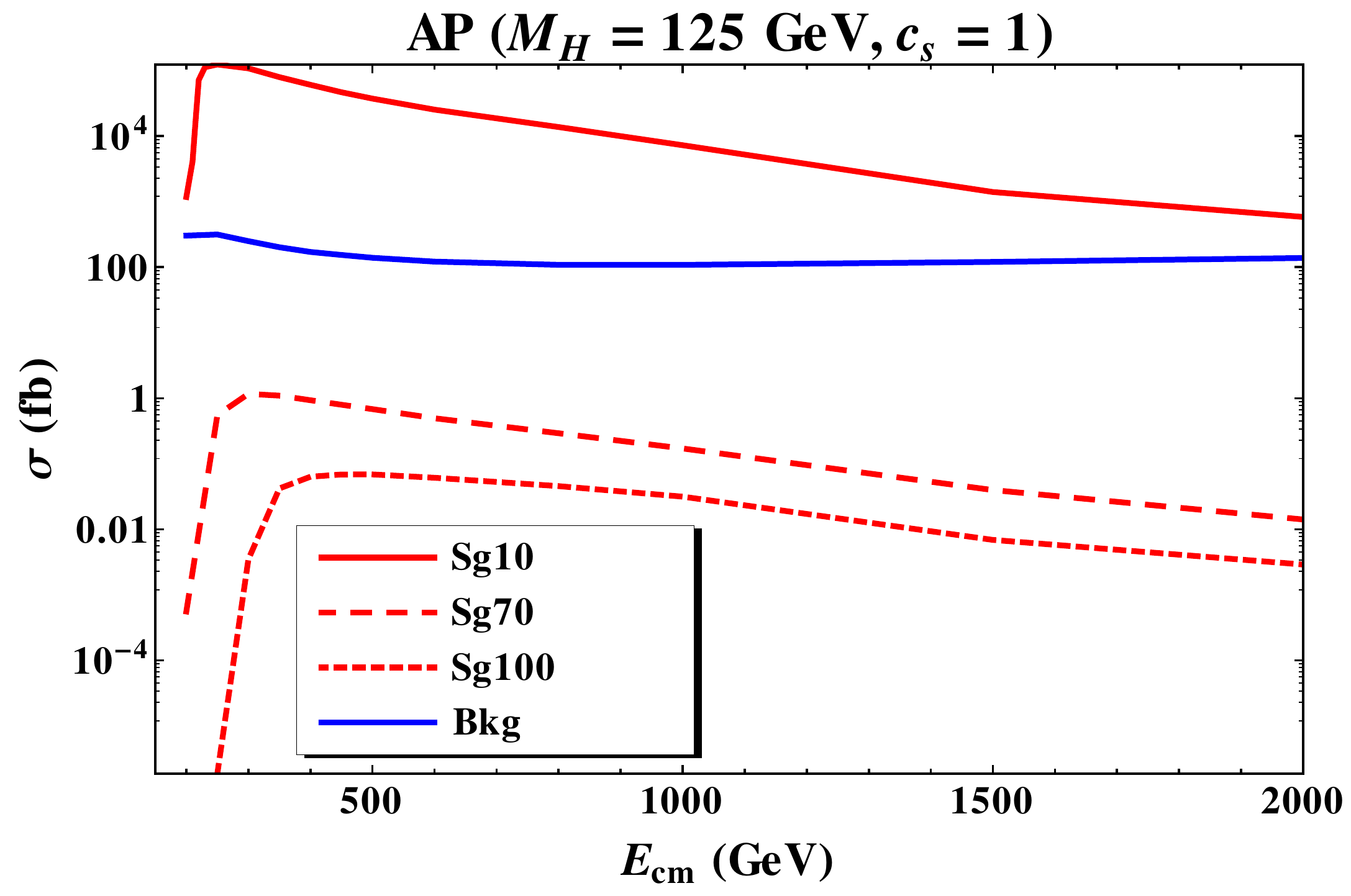}}\\
    \subfigure {\includegraphics[height=40mm,width=60mm]{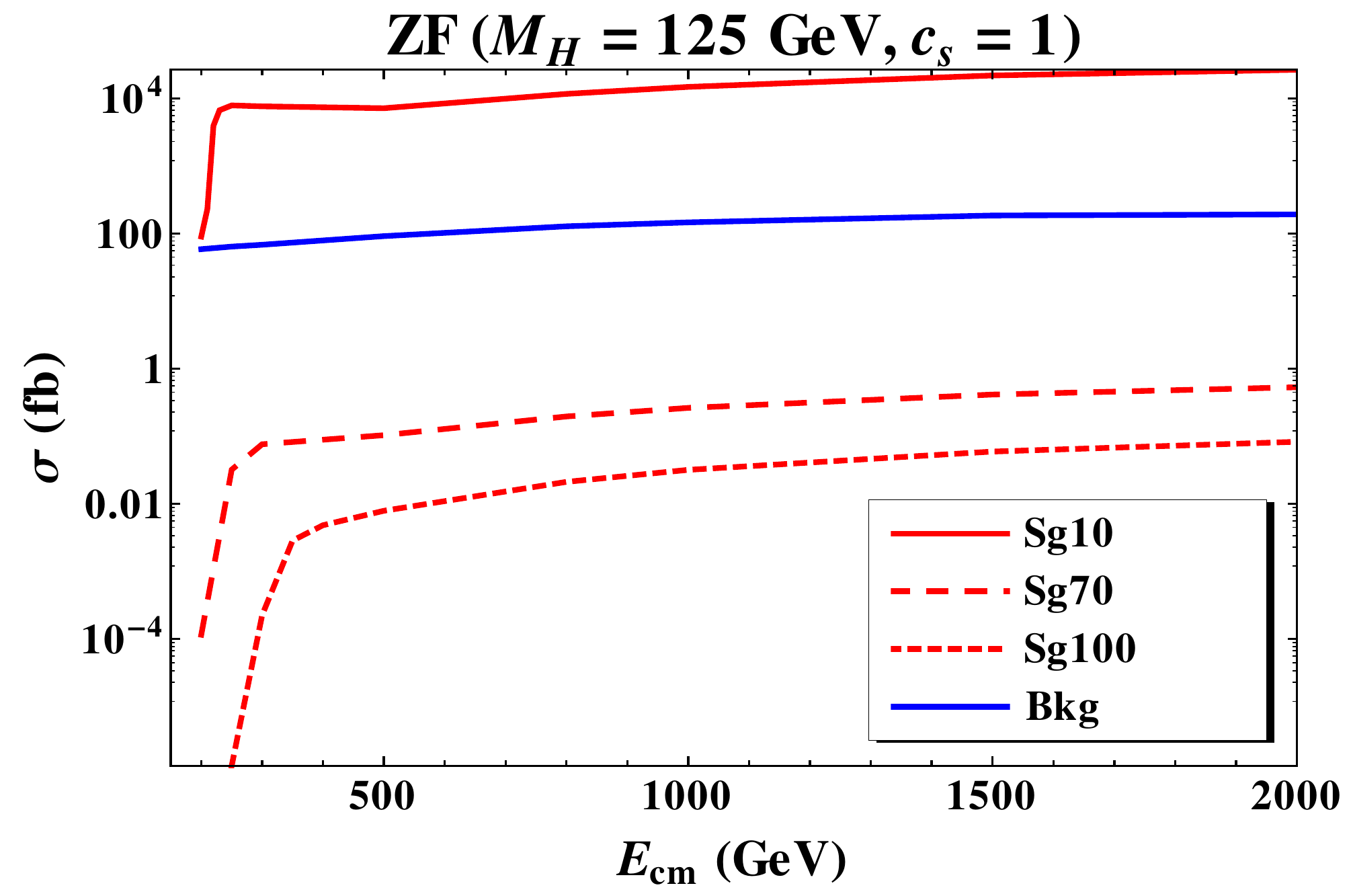}}  
  % }
  \caption{Signal (red) and background (blue) cross-section as functions of $E_{\rm CM}$, for AP and ZF channels, varying masses. For instance, legend ``Sg10'' indicates signal with $m_\phi=10$ GeV.}\label{fig:xsec_Ecm}
  \end{figure}

   Whether AP or ZF is the primary search channel depends on the center of mass energy ($E_{\rm CM}$) at which the collider runs. In Fig.(\ref{fig:xsec_Ecm}) we set $c_S=1$ and show the cross-sections of signal and background as a function of $E_{\rm CM}$ for the AP and ZF channels respectively. Mild basic kinematic cuts have been applied. A favorable polarization $P(e^-,e^+)=(+0.8,-0.5)$ has also been assumed. Such a high degree of polarization is only possible at a linear collider such as the ILC, but the general trend of energy dependence shown here also applies to a circular collider such as TLEP. 
    
   As can be seen from Fig.(\ref{fig:xsec_Ecm}), for $E_{\rm CM}\gtrsim500$ GeV, in the AP channel as $E_{\rm CM}$ increases, $\sigma_S^{\rm AP}$ falls as a power law, while $\sigma_B^{\rm AP}$ increases logarithmically. On the other hand, for the ZF channel both $\sigma_S^{\rm ZF}$ and $\sigma_B^{\rm ZF}$ increase logarithmically as $E_{\rm CM}$ increases. The logarithmic increases respect the Froissart unitarity bound~\cite{Froissart:1961ux}, and arise from exchange of particles with masses much less than $E_{\rm CM}$ in the t-channel~\cite{Chanowitz:1985hj}. As can be seen from Fig.(\ref{fig:xsec_Ecm}), the upshot is that AP constitutes the primary search channel for $\phi$ at $E_{\rm CM}\lesssim1$ TeV, while ZF becomes important for $E_{\rm CM}\gtrsim1$ TeV. Meanwhile, notice that in both channels $\sigma_S$ falls rapidly as $m_\phi$ increases beyond $m_H/2$, as a result of the structure of the intermediate $H^{*}$ propagator. As a consequence, even for $c_S$ of $O(1)$, our sensitivity to theories where $\phi$ is heavier than about 100 GeV is rather limited.
 
\section{Analysis}
\subsection{ILC}

\underline{\textbf{250 GeV run  at ILC}}\\
\indent At $E_{\rm CM}=250$ GeV, we focus on the case of $m_\phi<m_h/2$, since this energy is close to optimal for on-shell Higgs production, but the cross section for producing heavier $\phi$ through off-shell Higgs decay is small. The primary search channel is AP. We focus on signal events where the $Z$ decays hadronically which maximizes the signal cross-section, while $e^+e^-\rightarrow\nu\bar{\nu}+jj$ (mostly from $e^+e^-\rightarrow ZZ$) constitutes background. At a linear collider such as the ILC the initial $e^+e^-$ beams can be highly polarized, which is efficient in improving signal significance. We impose the polarizations $P(e^-,e^+)=(+0.8,-0.5)$, which are possible to realize in practice~\cite{Adolphsen:2013kya}. We apply this choice throughout our ILC studies, for different choices of $E_{CM}$. Note that this choice of polarizations differs in sign from that used in the examples given in section 1.2.8 of \cite{snowmass}, e.g.  $P(e^-,e^+)=(-0.8,+0.3)$. The reason for our choice here is that although the signs of the polarizations used in~\cite{snowmass} optimize the signal rate, they bring in an even larger increase in the background rate.  As a result, the significance is reduced by a factor of $\sim2$ compared to our choice.\\
\indent
    Here the most useful selection cut is on the missing invariant mass (MIM) in the event. Note that the MIM cut is only possible at a lepton collider since we have full information about the 4-momenta of the initial beams. In this case where $\phi$'s are produced from on-shell $H$ decays, we simply require MIM to reconstruct the Higgs mass peak around 125 GeV. On the other hand, the MIM of the background centers around $m_Z\approx90$ GeV, because the $\cancel{E}$ mostly comes from $Z\rightarrow\nu\bar{\nu}$. Since the MIM reconstruction depends on the energies of the two jets from $Z$ decay, in order to justify the robustness of the results from this parton level study we need to verify that the centers of the MIM distributions for signal and background are well separated even after taking into account the jet energy resolution of the detector. The jet energy resolution at the ILC can realistically be expected to be as good as $3\%$ in order to allow  $W$ and $Z$ bosons that decay hadronically to be distinguished \cite{Behnke:2013lya}. For a simple estimate, with $E_{CM}=250$ GeV, dijet energy from $Z$ decays is 110 GeV for the signal but 125 GeV for the background, which can be well separated with $3\%$ jet energy resolution. In addition, we require invariant dijet mass $M_{jj}$ to be around $m_Z$. With cuts $115{\rm~GeV} < {\rm MIM} < 135\rm~GeV$ and $70{\rm~GeV}< M_{jj} < 110\rm~GeV$, we found the $2\sigma$ reach for invisible Higgs BR to be $0.13\%$ with 1000 $\rm fb^{-1}$ data.\\

\begin{figure}
  % \mbox{   
   \subfigure   {\includegraphics[height=40mm,width=60mm]{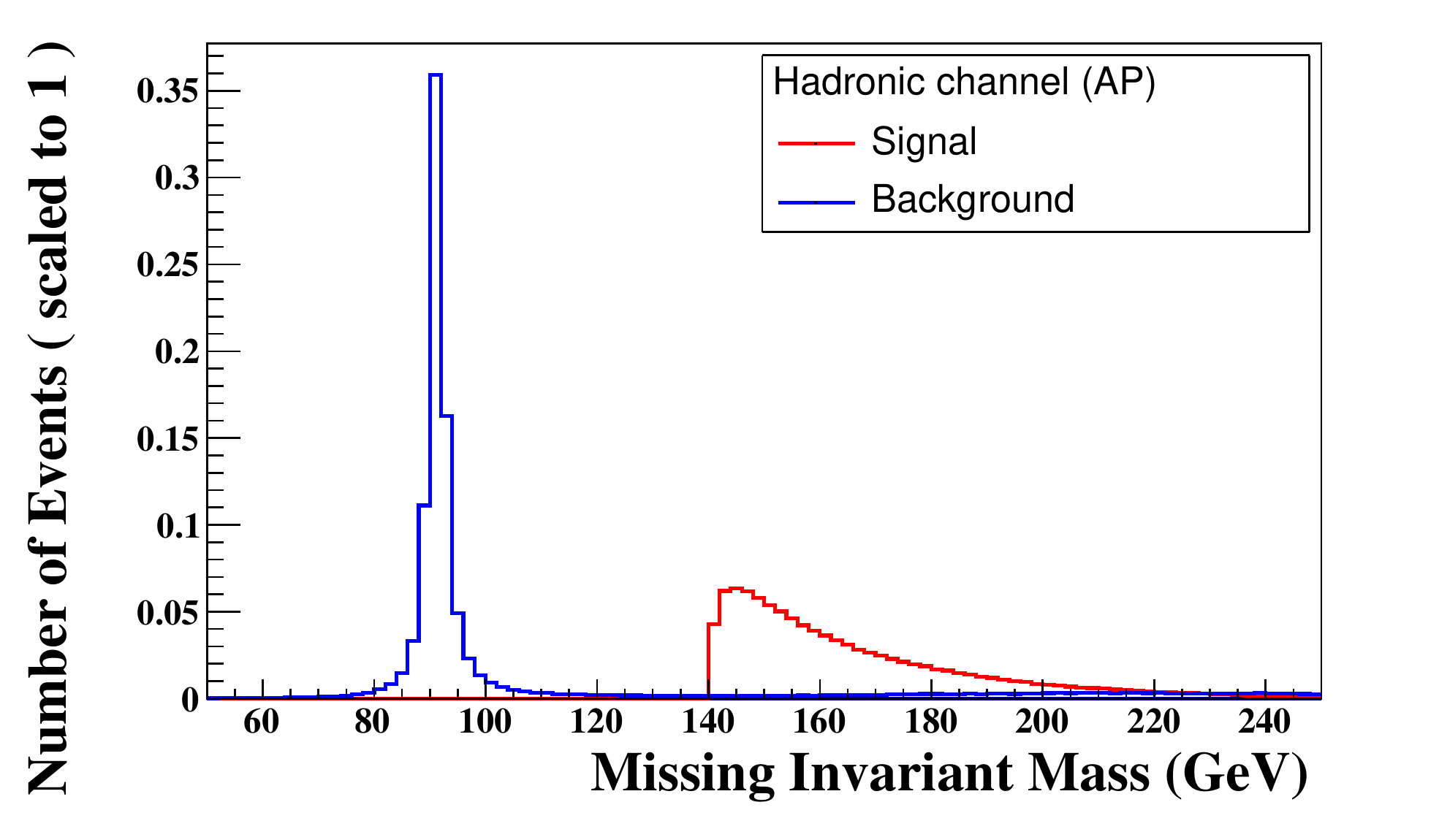}}\\
    \subfigure {\includegraphics[height=40mm,width=60mm]{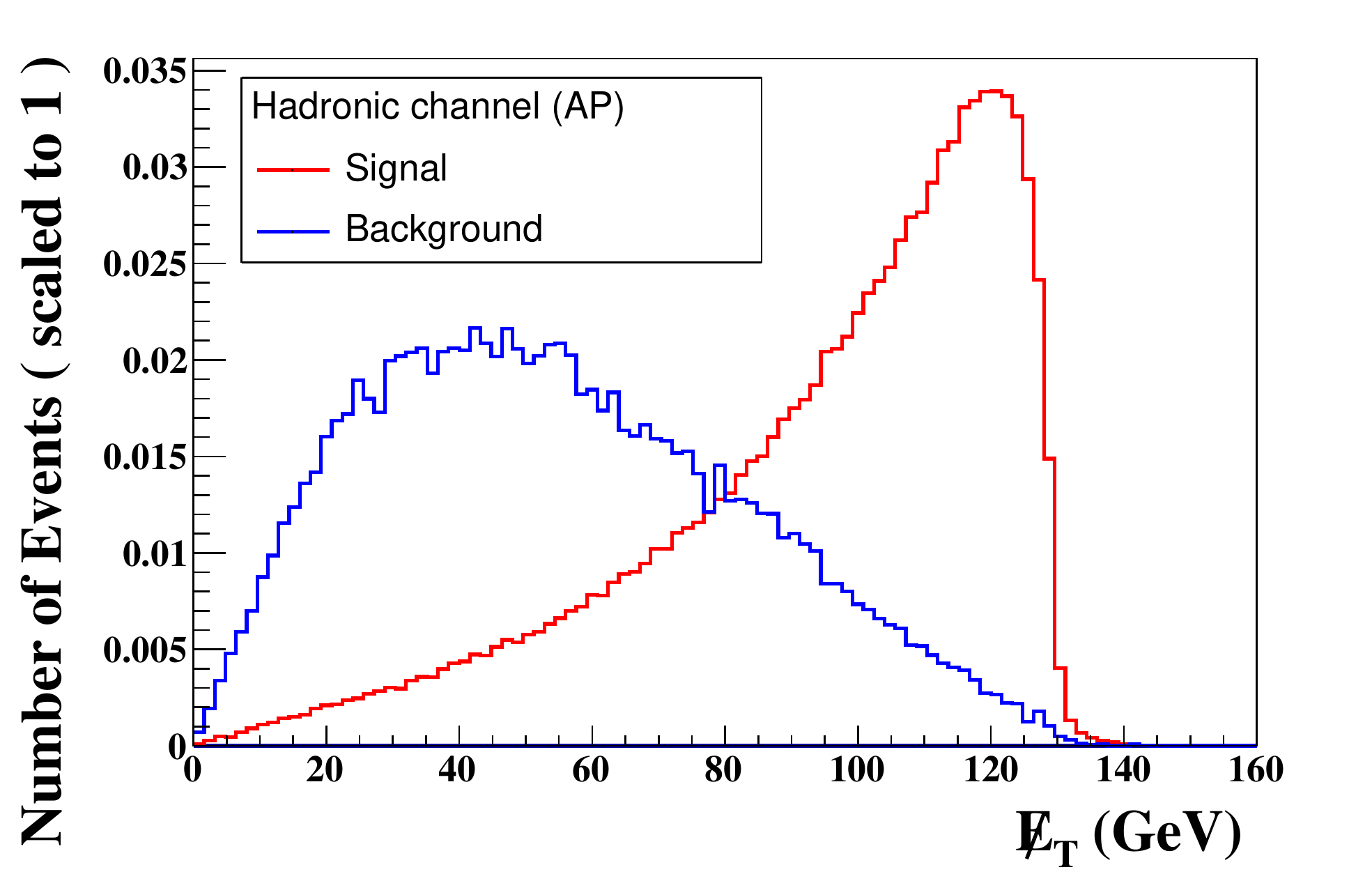}}  
   %}
  \caption{Signal vs. background distribution of MIM and $\cancel{E}_T$ variables ($E_{\rm CM}=350$ GeV, AP).}\label{fig:SvsB350}
  \end{figure}

\underline{\textbf{350 GeV run at ILC}}\\
\indent Next we consider the 350 GeV run at the ILC. The primary search channel is still AP. This larger $E_{CM}$ allows the study of heavier $\phi$ with $m_\phi>\frac{m_H}{2}$ produced through off-shell Higgs decay. In this case the MIM cut is still very efficient, although for a more subtle reason. The energy threshold for pair production of $\phi$ sets a lower limit on MIM around $2m_\phi$ for the signal events, which is separated from the narrow peak around $m_Z$ where the background is concentrated. Since the matrix element of the signal process falls rapidly as the $H^*$ propagator goes further off-shell, the MIM of the signal peaks slightly above $2m_\phi$. Considering the effects of detector resolution, for $m_\phi=70$ GeV the dijet energy from $Z$ decays is centered around 159 GeV for the signal and around 175 GeV for the major background, which can be separated with $3\%$ resolution. Therefore the application of MIM cut in this parton level study can be justified for $m_\phi\gtrsim70$ GeV. Another particularly useful selection cut is on $\cancel{E}_T$. The $\cancel{E}_T$ cut is useful because $\cancel{E}_T$ balances $E_T$ from the visible jets, which arise from the decay of a $Z$, and are more centrally produced in the signal process. For illustration, in Fig.(\ref{fig:SvsB350}) we show the signal vs. background distribution of the MIM and $\cancel{E}_T$ variables for $m_\phi=70$ GeV.
In Table~\ref{tab: 350ILC} we list all the cuts used that optimize the signal significance, and summarize the cross-sections/efficiencies of signals and backgrounds after each cut. The example shown is for $m_\phi=70$ GeV and $c_S = 1$.
\begin{table}
\begin{tabular}{l*{3}{c}c}
\hline
Cuts(GeV)                 & $S (\rm fb)$ & $B (\rm fb)$ & Effic.($S$) & Effic.($B$) \\
\hline
Initial(unpol.)           & 0.89     & 285      &             &  \\
\hline
Polarization(+0.8,-0.5)   & 1.09     & 203      &             &   \\
$E_{j_1,j_2} < 120$             & 0.96     & 93.4     & 88.7\%      & 46.1\% \\
$MIM > 140$               & 0.96     & 13.8     & 100\%       & 14.8\% \\
$\cancel{E}_T > 105$      & 0.44     & 0.75     & 46\%        & 5.4\% \\
$70 < M_{jj} < 110$       & 0.44     & 0.71     & 99.4\%      & 94.7\% \\
\hline
\end{tabular}
\caption{Cuts, $\sigma_S, \sigma_B$ and efficiencies after each cut for $m_\phi=$70 GeV at 350 GeV ILC (AP), $c_S=1$.\label{tab: 350ILC}}
\end{table}
We perform a similar analysis for $\phi$ mass of 80 GeV. With $\mathcal{L} = 1000 \rm ~fb^{-1}$ and $c_S = 1$, the reach at a 350 GeV ILC in terms of significance is summarized in the table below.

\begin{tabular}{c*{3}{c}c}
\hline
 AP ($c_S=1$,& 350 GeV,  $\mathcal{L} =$ & $ 1000 \rm ~fb^{-1}$) & & \\
\hline
$m_{\phi}$      & $70 \; \rm GeV$ & $80 \; \rm GeV$ &  &  \\
\hline
$S/ \sqrt{B}$   & 16.6        &   4.1   &     &  \\
\hline
\end{tabular}\\ \\

\underline{\textbf{500 GeV run at ILC}}\\
\indent At 500 GeV the primary search channel is still AP, and we continue to focus on the off-shell decay scenario. The MIM cut continues to be very efficient. Nonetheless, at $E_{CM}=500$ GeV, for $m_\phi=70$ GeV, the dijet energy from $Z$ decay is centered around 239 GeV for the signal and around 250 GeV for the major background, which can barely be separated with $3\%$ jet energy resolution. Therefore to ensure the robustness of this parton level analysis against detector smearing, here we limit our consideration to heavier $m_\phi$, starting from 80 GeV.
In addition to the MIM cut, a cut on the sum of $|p_T^i|$ of the visible jets, $H_T$, is found to be particularly useful. The $H_T$ cut is useful for the same reasons as the $\cancel{E}_T$ cut we discussed earlier. \\
\indent
We summarize all the cuts applied and their efficiencies on signals and backgrounds in Table~\ref{tab: 500ILC}, for the example with $m_\phi=80$ GeV and $c_S = 1$.
\begin{table}
\begin{tabular}{l*{3}{c}c}
\hline
Cuts(GeV)                 & $S (\rm fb)$ & $B (\rm fb)$ & Effic.($S$) & Effic.($B$) \\
\hline
Initial(unpol.)           & 0.20     & 348      &             &  \\
\hline
Polarization(+0.8,-0.5)   & 0.25     & 139.9    &             &   \\
$160 < MIM < 210$         & 0.17     & 2.3      & 66.1\%      & 1.6\% \\
$70 < M_{jj} < 110$       & 0.16     & 1.78     & 98.4\%      & 78.1\% \\
$H_T(jj) > 175$           & 0.12     & 0.52     & 74\%        & 29.3\% \\
$35 < p_{j_1,j_2} < 175$  & 0.10     & 0.38     & 86.2\%      & 72.7\% \\
\hline
\end{tabular}
\caption{Cuts, $\sigma_S, \sigma_B$ and efficiencies after each cut for $m_\phi=$80 GeV at 500 GeV ILC (AP), $c_S=1$.\label{tab: 500ILC}}
\end{table}
In the table below we summarize the reach at 500 GeV ILC with $\mathcal{L} = 1000 \rm ~fb^{-1}$ and $c_S = 1$, for varying masses.\\

\begin{tabular}{c*{3}{c}c}
\hline
 AP & ($c_S=1$,&500 GeV, & $\mathcal{L} = 1000 \rm ~fb^{-1}$)  & \\
\hline
$m_{\phi}$                     &  & $80 \; \rm GeV$ & $90 \; \rm GeV$ & $100 \; \rm GeV$ \\
\hline
$S/ \sqrt{B}$   &     & 5.4         & 2.5         & 1.2 \\
\hline
\end{tabular}\\ \\

\begin{figure}
%   \mbox{   
   \subfigure  {\includegraphics[height=40mm,width=60mm]{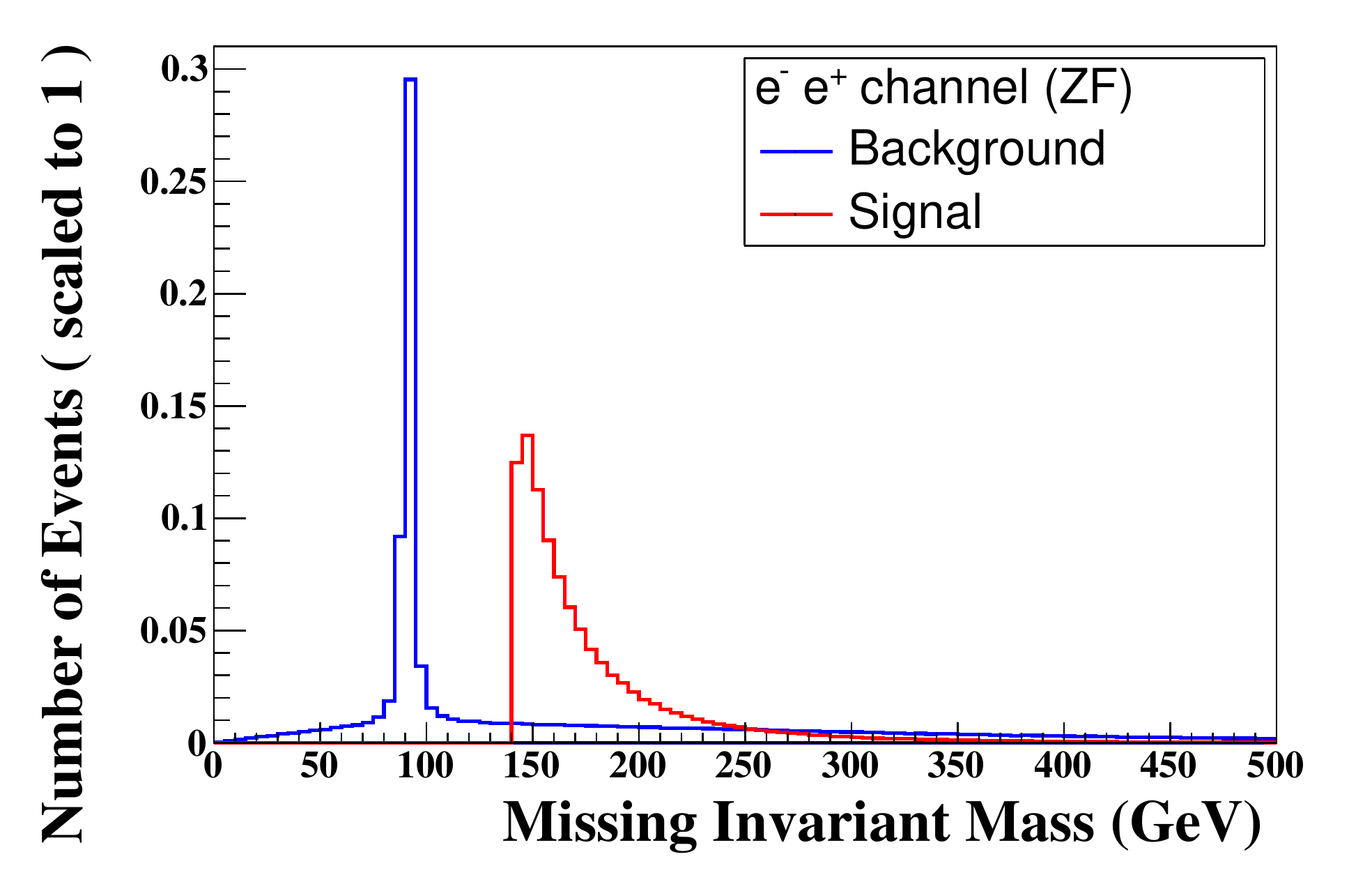}}\\
    \subfigure {\includegraphics[height=40mm,width=60mm]{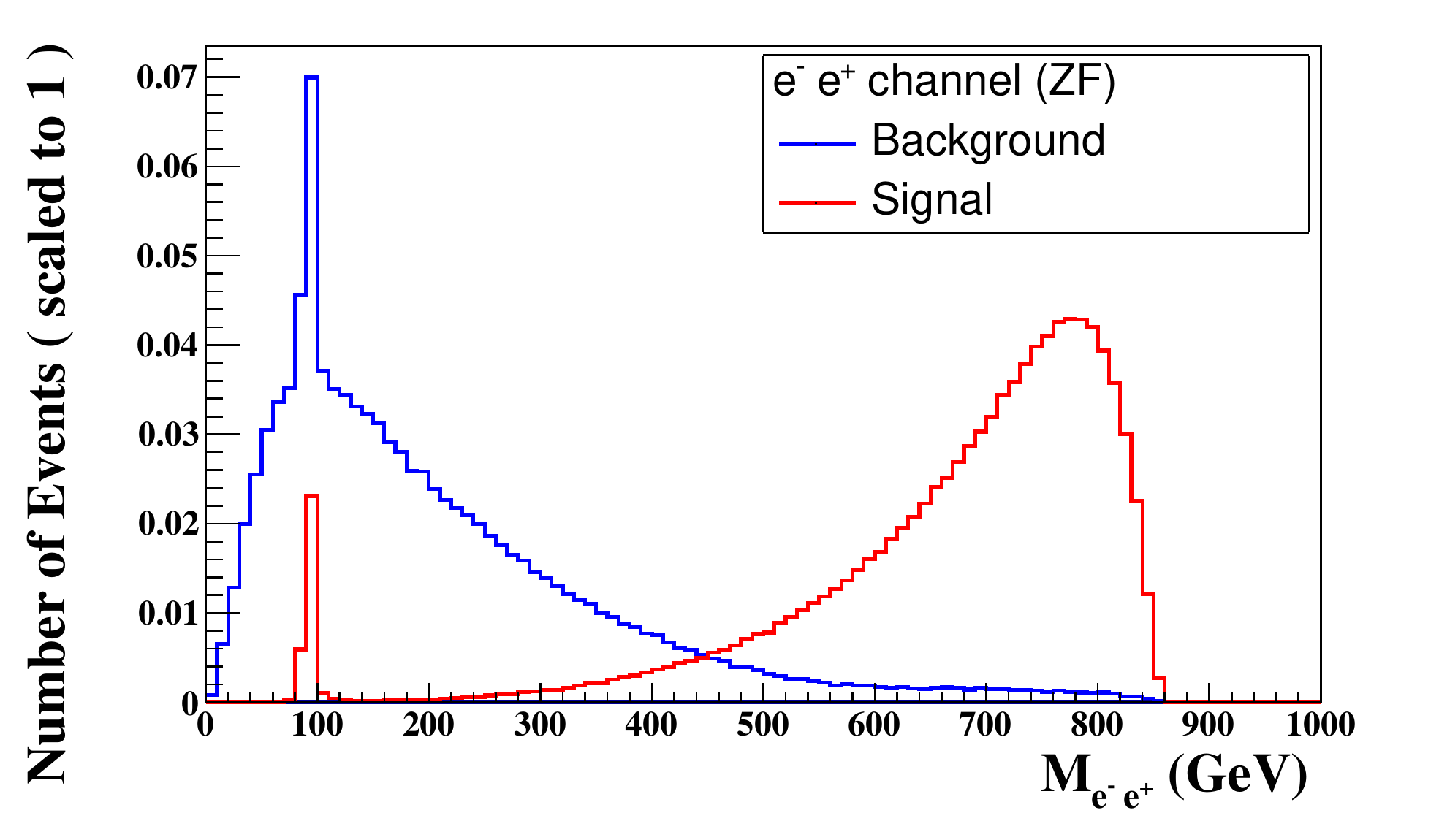}}  
  % }
  \caption{Signal vs. background distribution of MIM and $M_{e^{-} e^{+}}$ variables ($E_{\rm CM}=1$ TeV, ZF).}\label{fig:SvsB1TeV}
  \end{figure}

\underline{\textbf{1 TeV run at ILC}}\\
\indent
 At 1 TeV, the ZF channel becomes significant. Meanwhile, for the 
hadronic AP channel, due to the high energies of the jets from $Z$ 
decays at $E_{CM}=1$ TeV, after detector smearing the MIM distributions 
for signal and background may not be well separated even if $m_\phi$ is 
as large as 100 GeV. Therefore for a reliable parton level study, we 
focus on the ZF channel. Since the only visible final states are 
leptons, the detector energy resolution is not expected to be a major 
source of error. For the ZF channel the background events involve 
$e^+e^-\rightarrow e^+e^-\nu\bar{\nu}$. MIM continues to be a powerful 
cut here. The MIM peak for signal is narrower than in the AP case due to 
the extra phase space suppression on $\phi$, which is a consequence of 
the soft/collinear enhancement in $Z^*$ emission. In addition, useful 
cuts include invariant mass of the $e^+e^-$ pair, $H_T(e^+e^-)$ and 
$\cancel{E}_T$. These variables capture the feature that in the ZF 
signal events the outgoing $e^+, e^-$ tend to lie along the 
forward/backward directions, while the missing energy is more central. 
Fig.(\ref{fig:SvsB1TeV}) illustrates S~vs.~B distributions of some of 
the key variables.
With $m_\phi=70$ GeV and $c_S = 1$ as an example, in Table~\ref{tab:1000ILC} we list the selection cuts for the ZF channel, and the cross-sections/efficiencies of signal and background after each cut.
\begin{table}
\begin{tabular}{l*{3}{c}c}
\hline 
 & ZF & & & \\
\hline
Cuts (GeV)                   & $S (\rm fb)$ & $B (\rm fb)$ & Effic.($S$) & Effic.($B$) \\
\hline
Initial(unpol.)              & 0.31  & 456.4    &             &   \\
\hline
Polarization(+0.8,-0.5)         & 0.26    & 148      &             &   \\
$140 < MIM < 175$            & 0.17    & 8.42     & 64.7 \%     & 5.65 \% \\
$M_{e^{+} e^{-}} > 700$      & 0.10   & 0.38    &  61.2 \%    & 4.58 \%  \\
$H_T(e^+e^-) < 260$          & 0.099   & 0.26   & 94.1 \%     & 66.3 \% \\
$\cancel{E}_T > 70$          & 0.061   & 0.053   &  62.1 \%    & 20.9 \%   \\
\hline
\end{tabular}
\caption{Cuts, $\sigma_S, \sigma_B$ and efficiencies after each cut for $m_\phi=$70 GeV at 1 TeV ILC ( ZF), $c_S=1$.\label{tab:1000ILC}}
\end{table}
 In the table below we summarize the reach at 1 TeV with $\mathcal{L} = 
1000 \rm ~fb^{-1}$ and $c_S = 1$ for different $\phi$ masses in the ZF 
channel. \\ 

\begin{tabular}{c*{3}{c}c}
 \hline
& ZF & ($c_S=1$, &1 TeV, & $\mathcal{L} = 1000 \rm ~fb^{-1}$)   \\
\hline
$m_{\phi}$                     & $70 \; \rm GeV$ & $80 \; \rm GeV$ & $90 \; \rm GeV$ & $100 \; \rm GeV$ \\
\hline
$S/ \sqrt{B}$   & 8.4        & 2.7         & 1.4         & 0.8 \\
\hline
\end{tabular}

A careful analysis of the ZF channel may allow the spin and mass of the 
DM to be determined~\cite{Andersen:2013rda}. We leave this for future work.

\subsection{TLEP}
\underline{\textbf{240 GeV run at TLEP}}\\
\indent
As in the  case of the 250 GeV run at the ILC, we focus on the case with on-shell Higgs decay in the AP channel. At a circular collider like TLEP, the initial beams cannot be highly polarized in the preferred way at the CM energies of interest~\cite{Gomez-Ceballos:2013zzn}. Therefore the potentially efficient polarization selection does not apply here. The jet energy resolution for TLEP is still uncertain at this point~\cite{Gomez-Ceballos:2013zzn}. Here we again assume an optimal case with about $3\%$ resolution, which allows hadronically decaying W and Z to be separated, 
and justifies the reliability of the MIM cuts used in our parton level analysis. After applying $115 \rm~GeV< MIM < 135\rm~GeV$ cut, we find the best $2\sigma$ reach for invisible Higgs BR at TLEP to be $0.08\%$, with $10^4$ fb$^{-1}$ data.\\

\underline{\textbf{350 GeV run at TLEP}}\\
\indent This analysis is similar to the  case of 350 GeV at the ILC, and we again focus on the case with off-shell Higgs decays in the AP channel. As before we take $m_\phi=70$ GeV, $c_S = 1$ as an example. Table~\ref{tab: 350TLEP} lists the imposed selection cuts as well as the cross sections/efficiencies of signals and backgrounds after each cut.
\begin{table}
\begin{tabular}{l*{3}{c}c}
\hline
Cuts (GeV)             & $S (\rm fb)$ & $B (\rm fb)$ & Effic.($S$) & Effic.($B$) \\
\hline
Initial(unpol.)        &  0.89   & 285.0     &              &  \\
$E_{j_1,j_2} < 130$    &  0.86   & 204       & 97\%         & 71.3\% \\
$ MIM > 135$           &  0.86   & 96.4      & 100 \%       & 47.3 \% \\
$70 < M_{jj} < 110$    &  0.84   & 72.6      & 98 \%        & 75.3 \% \\
$\cancel{E}_T > 100$   &  0.45   & 4.9       & 53.9 \%      & 6.8 \%  \\
\hline
\end{tabular}
\caption{Cuts, $\sigma_S, \sigma_B$ and efficiencies after each cut for $m_\phi=$70 GeV at 350 GeV TLEP (AP), $c_S=1$.\label{tab: 350TLEP}}
\end{table}
The reach for different $m_\phi$ is summarized below.\\

\begin{tabular}{c*{2}{c}c}
\hline
AP & ($c_S=1$, &350 GeV, & $\mathcal{L} = 2500 \rm ~fb^{-1}$)   \\
\hline
$m_{\phi}$             & $70 \; \rm GeV$ & $80 \; \rm GeV$ & $90 \; \rm GeV$ \\
\hline
($S/ \sqrt{B}$)    & 11.7         & 2.5         & 0.8   \\
\hline
\end{tabular}\\ \\ 

\subsection{Combined results}
\begin{figure}
\begin{centering}
    \includegraphics[height=70mm,width=85mm]{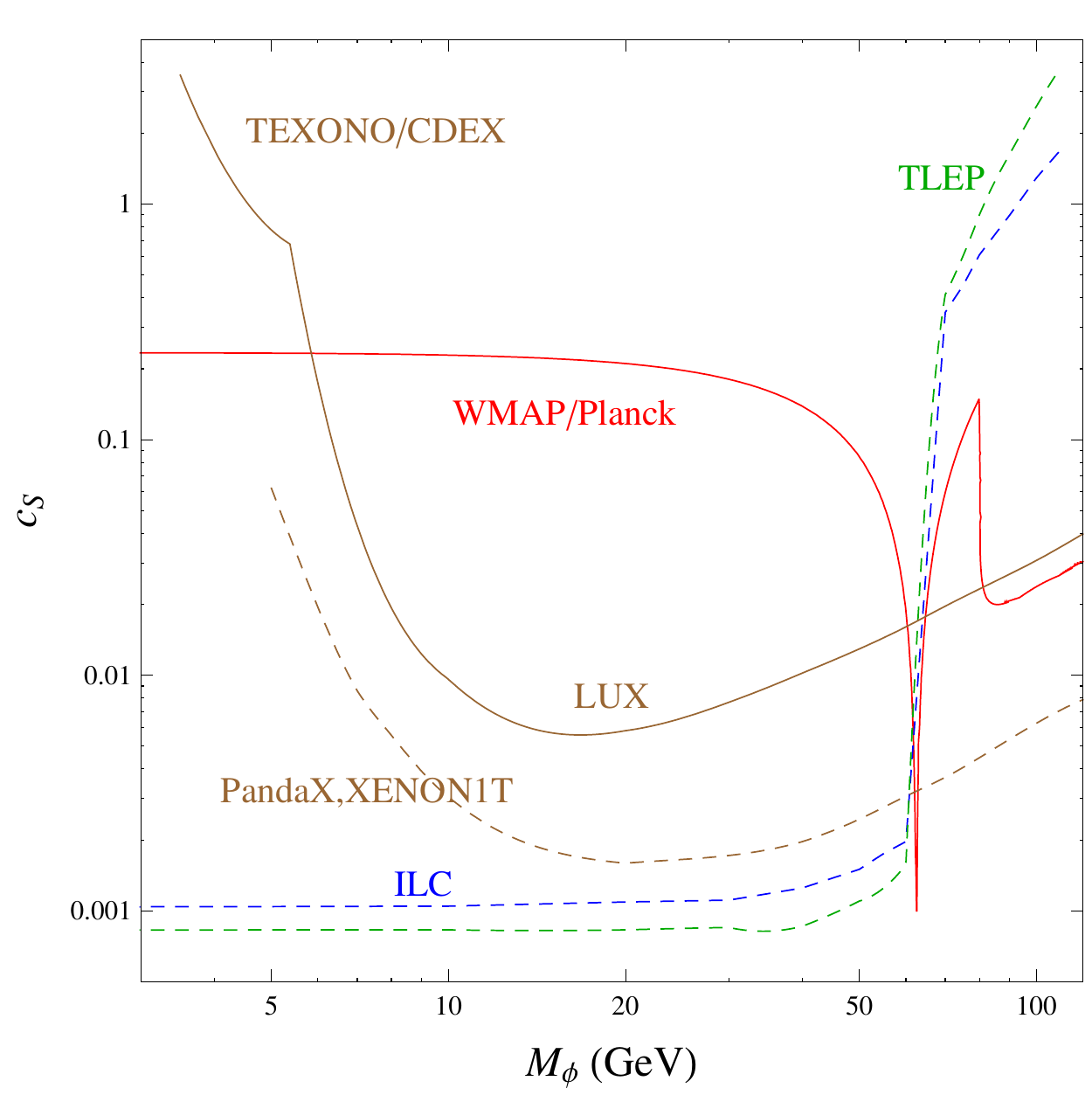}
  \caption{$2\sigma$ significance reach contours for searches at the ILC (dashed blue) and TLEP (dashed green). We assume a jet energy resolution of $3\%$ for both ILC and TLEP.  Other details about getting the ILC and TLEP contours are explained in the main text. $90\%$ exclusion region by current best limits, from LUX and TEXONO/CDEX (solid brown). $90\%$ projected exclusion limit by next generation experiments PandaX, XENON1T (dashed brown). Region consistent with WMAP/Planck data (red).}\label{fig:combined}
  \end{centering}
  \end{figure}

Now we present the sensitivity contour plots, showing the 2$\sigma$ limits on this scenario as a function of $m_\phi$ and $c_S$ in Fig.(\ref{fig:combined}). The limits shown for the on-shell
case are obtained from 1000 \rm fb$^{-1}$ at 250 GeV for ILC, and from 10000 \rm fb$^{-1}$ at 240 GeV for TLEP. The limits on the $m_\phi$ = 70 GeV case are obtained from 1000 \rm fb$^{-1}$ at 350 GeV for  the ILC, and from 2500 \rm fb$^{-1}$ at 350 GeV for TLEP.  For larger $m_\phi$, the ILC results are obtained from 1000 \rm fb$^{-1}$ at 500 GeV and from 2500 \rm fb$^{-1}$ at 350 GeV for TLEP. We have assumed a $3\%$ jet energy resolution for both ILC and TLEP. As mentioned earlier, this energy resolution has been shown to be feasible for the ILC, but is yet to be validated for TLEP. To illustrate the particularly interesting case where $\phi$ is DM, we also display the parameters that give rise to the observed relic abundance from WMAP/Planck, current best limits from DM direct detection experiments~\cite{Akerib:2013tjd,Zhao:2013xsf,Li:2013fla}, as well as the projected reach at the next generation DM experiments~\cite{Aprile:2012zx,pandax}. As can be seen from Fig.(\ref{fig:combined}), future $e^+e^-$ colliders are highly complementary to DM direct detection experiments. In particular, in the region of light DM masses where direct detection experiments are weakest, the
collider limits can be very strong. More generally, for $m_\phi<m_h/2$, linear colliders can bound $c_S$ below $\sim10^{-3}$ (corresponding to invisible BR $\lesssim 0.2\%$), which is even better 
than the the next generation DM experiments with highest sensitivity, such as XENON1T and PandaX. 

\section{Conclusion}
In this work we have demonstrated that the next generation $e^+e^-$ colliders offer an excellent opportunity to explore a hidden sector through the Higgs portal. With the minimal singlet scalar model as an example, in the light mass range where $m_\phi<m_h/2$, the sensitivity at $e^+e^-$ colliders is superior to that of the next generation DM direct detection experiments. Even in the challenging case when $m_\phi>m_h/2$, scalar masses as high as $m_\phi = 100$ GeV can be probed if the couplings of $\phi$ to the Higgs are of $O(1)$ strength. Furthermore, in 
the event of a discovery at a DM direct detection experiment, an $e^+e^-$ collider may be able to provide clear evidence of the underlying Higgs portal by studying the distribution of the MIM variable. The DM direct detection experiment would not be able to disentangle such details. These conclusions are based on a parton level analysis, and a more dedicated study that includes initial state radiation, hadronization and detector effects is warranted. This is left for future work. 

\section*{Acknowledgements}
We thank Andrei Gritsan and Rick van Kooten for helpful discussions. The 
authors are supported in part by NSF grant PHY-0968854.  S.~Hong is 
supported in part by a fellowship from The Kwanjeong Educational 
Foundation.

\end{document}